\documentclass[prd,twocolumn,a4paper,superscriptaddress,floatfix]{revtex4}
\usepackage{graphicx}
\usepackage{amssymb}
\usepackage{amsfonts}
\usepackage{amsmath}
\begin{document}

\newcommand{\be}{\begin{equation}}
\newcommand{\ee}{\end{equation}}
\newcommand{\bq}{\begin{eqnarray}}
\newcommand{\eq}{\end{eqnarray}}
\newcommand{\bsq}{\begin{subequations}}
\newcommand{\esq}{\end{subequations}}
\newcommand{\bc}{\begin{center}}
\newcommand{\ec}{\end{center}}
\newcommand {\R}{{\mathcal R}}
\newcommand{\al}{\alpha}
\newcommand\lsim{\mathrel{\rlap{\lower4pt\hbox{\hskip1pt$\sim$}} \raise1pt\hbox{$<$}}}
\newcommand\gsim{\mathrel{\rlap{\lower4pt\hbox{\hskip1pt$\sim$}} \raise1pt\hbox{$>$}}}

\title{ 
AdS black disk model for  small-x DIS}

\author{Lorenzo Cornalba, Miguel S. Costa}
\affiliation{Centro de F\'{\i}sica do Porto e Departamento de F\'{\i}sica e Astronomia da Faculdade de Ci\^encias da Universidade do Porto, Rua do Campo Alegre 687, 4169-007 Porto, Portugal}
\author{Jo\~ao Penedones}
\affiliation{Kavli Institute for Theoretical Physics, University of California, Santa Barbara, CA 93106-4030, USA}

\begin{abstract}
Using the approximate conformal invariance of QCD at high energies
we consider a simple AdS  black disk model to describe
saturation in DIS.
Deep inside saturation the 
structure functions have
the same power law scaling,
$F_T \sim F_L \sim  x^{-\omega}$, where 
$\omega$ is related to the expansion rate of the black disk with energy. Furthermore,
the  ratio $F_L/F_T$  is given by the universal value $\frac{1+\omega}{3+\omega}$, independently of the target.
For  $\gamma^*-\gamma^*$
scattering at high energies we obtain explicit expressions and ratios
for the total cross sections of transverse and longitudinal  photons in terms of the single parameter $\omega$.
\end{abstract}
\maketitle

\section{\label{sint}Introduction}

Deep inelastic scattering (DIS) has been of primary importance for understanding the partonic  structure of hadrons. A classical example is the 
Callan-Gross relation $$ F_2 = 2x F_1\,,$$
 which holds in the  limit of large photon virtuality $Q$
at fixed  Bjorken $x$, and shows that quarks have spin 1/2 \cite{CallanGross}.

In this Letter we shall focus on the   kinematical limit of   fixed $Q$ and  $x\ll 1$,
which corresponds to the Regge limit of large center of mass energy, $s\approx  Q^2/x$, with other kinematical invariants fixed. 
In this kinematical window one observes a power like growth in $1/x$ of the hadron structure functions \cite{sat}.
The hadron becomes a dense gluon
medium so that the picture of the hadron made of weakly interacting partons is no longer valid. 
Although the coupling $\alpha_s$ may still be small, say for hard probes with 
 $Q  \gsim 1 \,{\rm GeV}$,
to understand low $x$ structure functions one needs to include diagrams such as those of  order 
$[\alpha_s\ln(1/x)]^n$ due to the kinematical enhancement. Thus, for hard probes, low $x$ DIS is the ideal 
ground to explore the approximate conformal symmetry of QCD in a 
situation where the gluon density  inside the hadron is so high that nonlinear effects are nevertheless important.

A new approach to DIS using the AdS/CFT  duality \cite{AdSCFT} was put forward in \cite{PS},
and further explored in [5-19].
We shall consider the AdS/CFT approach to saturation in DIS \cite{HIM,Saturation}, and
model the saturated hadron with an AdS black disk (a black disk in conformal QCD), extending
the results of  \cite{Saturation} by including the polarization of the hard probe \cite{CCP09}. 
We will show that deep inside saturation the AdS black disk model
predicts  the scaling law
$F_L \sim F_T \sim x^{-\omega}$, with the universal ratio
\begin{align}
\frac{F_L }{F_T}\approx \frac{F_2 -2xF_1 }{2xF_1}
\approx   
\frac{1+\omega}{3+\omega}\,,
\label{Ratio}
\end{align}
where the constant $\omega$, as explained below, 
is related to the expansion rate of the AdS black disk with energy, independently of the
details of the target hadron,
We propose that this is a good description deep inside saturation, 
for $   1 \,{\rm GeV}  \lsim Q\ll Q_s$ where $Q_s$ is the saturation scale.
We also consider $\gamma^*-\gamma^*$ scattering,
obtaining universal ratios for the polarized cross sections.

The direct  measurement of the longitudinal structure function  $F_L$ is an important test 
to QCD.
For low $x$ such measurements  were  done at HERA,  still outside the saturation region \cite{Hera}.  
Other data from $\gamma^*-\gamma^*$ scattering 
at LEP2 \cite{DELPHI} are also outside this regime.
Future experiments, such as LHeC \cite{LHeC} or ILC \cite{ILC}, will enter the saturated domain, and will be able to test the model here
proposed.


\section{AdS Black Disk}

To define the scattering amplitude of an off-shell photon by a scalar target in a conformal field theory,
we consider the momentum space correlation function,
\begin{equation*}
(2\pi)^4\, \delta\left( \sum k_j \right) i\, T^{ab}(k_j) =  
\left\langle j^a(k_1) {\cal O}(k_2) j^b(k_3) {\cal O}(k_4) \right\rangle\,,
\end{equation*}
involving  a conserved current $j^a$ and a 
scalar primary operator  ${\cal O}$ of dimension $\Delta$.
We shall be interested
in the limit of large $s=-(k_1+k_2)^2$ with fixed momentum transfer $t= - (k_1+ k_3)^2 = -q_\perp^2$ and virtualities $k_i^2$.
As shown in \cite{CCP06,Saturation,CCP09}, 
this limit is conveniently described by the impact parameter representation
\begin{align}
 T^{ab}(k_j)  &\approx  2 i s  \int dl_\perp\,
 e^{iq_\perp\cdot l_\perp} 
 \label{IPrep}
 \\
 &\times \int \frac{dr}{r^3}\, \frac{d\bar{r}}{\bar{r}^3} \,
\Psi^{ab\ \tau}_{\ \ \mu}(r)\ \Phi (\bar{r})  \,
\left[ 1-e^{i\chi (S,L)}  \right]_{\ \tau}^\mu \,,
 \nonumber
\end{align}
where the phase shift $\chi_{\ \tau}^\mu$ is a tensor  that encodes all the dynamical information and depends on 
$$
S=r\bar{r} s \ ,\ \ \ \ \ \ \ \ \ \ 
\cosh L=\frac{r^2+\bar{r}^2+l_\perp^2}{2r\bar{r}}\,.
$$
The scalar function $\Phi$ is associated with the operator ${\cal O}$ and the 
tensor function $\Psi^{ab\ \tau}_{\ \ \mu}$   with the current operator $j^a$. Their explicit form 
was given in \cite{CCP09}.
It is important to note that this representation 
is valid for any value of the coupling, 
since it relies
only on conformal invariance.

The above conformal representation is quite natural from the view point of the dual AdS scattering process, with
transverse space given by the three-dimensional hyperbolic space $H_3$, whose boundary is the 
physical transverse  space $\mathbb{R}^2$. Using Poincar\'e coordinates,
$$
ds^2(H_3)=\frac{dr^2+ds^2(\mathbb{R}^2)}{r^2}\ ,
$$
we identify $L$ with the geodesic distance between two points that are separated by $l_\perp$ 
along $\mathbb{R}^2$ and have radial coordinates $r$ and $\bar{r}$.
The variable $S$ measures the local energy squared of the scattering process in AdS.
The Greek indices $\mu$ and $\tau$ in (\ref{IPrep}) label tangent directions to $H_3$, 
which are the physical polarizations of the AdS gauge field dual to the conserved current $j^a$. 
The functions 
\begin{equation*}
\Psi^{ab\mu\tau} (r)=  
\psi_{in}^{a\mu}(r)\,\psi_{out}^{b\tau}(r)\,,\ \ \ \ \ \ 
\Phi (\bar{r})= \phi_{in}(\bar{r})\,\phi_{out}(\bar{r})   \,,
\end{equation*}
are given by the product of the radial part of the incoming and outgoing 
dual AdS  fields.
These 
functions are non-normalizable because they are produced by a plane wave
source created by the dual operator at the boundary.

 We shall consider a  black disk model defined by a phase shift in the impact parameter representation (\ref{IPrep}) given by
$$
\left[1- e^{i\chi(S,L)}\right]^\mu_{\ \tau}=\Theta\big(L_s(S)-L\big)\,\delta^\mu_{\ \tau}\ ,
$$
where the radius $L_s$ of the disk increases with energy  as
\begin{equation}
L_s(S) \approx \omega \log S\ .
\label{AdSdisk}
\end{equation}
Note that the size of the disk is independent of the  dual AdS gauge field polarization,  so that this simple model is characterized by the single parameter $\omega>0$.
To motivate this model,
in section \ref{Geometric Scaling} we assume Reggeon exchange and consider the limiting cases corresponding to
the BFKL Pomeron at weak coupling and 
$\mathcal{N}=4$ SYM at strong coupling. We then use geometric scaling observed in DIS at low $x$ to phenomenologically fix  $\omega$.

\section{Deep Inelastic Scattering}

The total DIS cross section, and corresponding hadron structure functions, are related to 
the hadronic tensor
\begin{equation*}
  W^{ab}(k_j) =  i \int d^4y\,e^{i k_1\cdot y}
 \langle  k_2 | {\rm T} \left\{j^a(y)   j^b(0)\right\}  | k_2  \rangle\,,
\end{equation*}
where $j^a$ is the electromagnetic current and 
$| k_2  \rangle$ is the 
target hadron state of momentum $k_2$.
We define  the virtuality $Q^2=k_1^2  $, target mass $M^2=-k_2^2 $ and Bjorken 
$$
x=-\frac{Q^2}{2 k_1\cdot k_2} \approx 
 \frac{Q^2}{s}\,.
$$
Lorentz invariance and conservation restricts  $W^{ab}$ to 
$$
W^{ab}= \left( \eta^{ab}-\frac{k_1^a k_1^b}{k_1^2}\right)\Pi_1 
+\frac{2x}{Q^2}  \left( k_2^a+\frac{k_1^a  }{2x}\right)\left( k_2^b+\frac{k_1^b }{2{x}}\right) \Pi_2 \,.
$$
The structure functions $F_i$ satisfy $2\pi F_i={\rm Im}\, \Pi_i$.

At zero momentum transfer we can use the representation (\ref{IPrep}) to write the hadronic tensor as 
\begin{align}
W^{ab}  &\approx  4\pi i s
\int \frac{dr}{r^2}\, \frac{d\bar{r}}{\bar{r}^2} \,
\Psi^{ab\ \tau}_{\ \ \mu}(r)\ \Phi (\bar{r}) 
\nonumber \\
&\times \int_{\left| \ln \bar{r}/r \right|}^{+\infty} dL \sinh L \left[  e^{i \chi (S,L)}-1\right]^{ \mu }_{\ \tau }\,,
 \label{forward} 
\end{align}
where we did the angular integral in the impact parameter $l_{\perp}$ and traded $|l_{\perp}|$ in the
radial integration for the  AdS impact parameter $L$. 
Note that $\Phi(\bar{r})=|\phi(\bar{r})|^2$, where now  $\phi(\bar{r})$ is the radial part of the normalizable AdS wave function dual to the 
state $| k_2  \rangle$.
This wave function is localized in the IR around $\bar{r}\sim 1/M$.
Its explicit form in the IR region, where space is no longer AdS, will not be important in what follows, because we shall consider a hard
probe localized near the AdS boundary. 

The black disk model of the previous section gives
\begin{align}
W^{ab} & \approx  -2\pi i s
\int \frac{dr}{r^2}\, \frac{d\bar{r}}{\bar{r}^2} \,
\Psi^{ab\ \mu}_{\ \ \mu}(r)\ \Phi (\bar{r}) \nonumber \\
& \times
\left[ (sr\bar{r})^\omega + (sr\bar{r})^{-\omega}  - \frac{r}{\bar{r}} - \frac{\bar{r}}{r}  \right] \,.
\label{saturatedW}
\end{align}
At very low $x$ the first term dominates and we have
\begin{align}
W^{ab} & \approx  -2\pi i s^{1+\omega}
\int \frac{dr}{r^{2-\omega}} \,
\Psi^{ab\ \mu}_{\ \ \mu}(r)\,
\int \frac{d\bar{r}}{\bar{r}^{2-\omega}}\,\Phi (\bar{r})\,.
\label{lowxW}
\end{align}
To give the explicit form of $\Psi$ it is convenient to write
 \begin{align}
k_1 = \left(\sqrt{s} ,- \frac{Q^2}{\sqrt{s}},0\right)\ ,    \ \ \ \ \ \ \ 
k_2 = \left(\frac{M^2}{\sqrt{s}},\sqrt{s},0\right)\ ,
\label{k1k2}
\end{align}
 in light-cone coordinates $(+,-,\perp)$. 
 Then, following \cite{CCP09},  
  \begin{align}
\Psi^{ab\ \mu}_{\ \ \mu}(r) &
= -\frac{    \pi^{2} }{6  } C \, r^2  \int_0^\infty dudv \,
e^{-u-v-\frac{Q^2 r^2}{4u}-\frac{Q^2 r^2}{4v}}    
\nonumber\\ 
& \times\  
\left(
\begin{array}{ccc}
  \frac{ s r^2}{4uv }  &  \frac{v-1}{u}   
  &0\\
  \frac{u-1}{v} &   \frac{4(u-1)(v-1)}{s r^2}  &  0  \\
  0&0&\mathbb{I}  
 \end{array}
\right)\,,
\label{Psi}
\end{align}
where the matrix elements are also ordered by the light-cone coordinates.
In particular, we have
\begin{align*}
\Psi_{\ \ \mu}^{ij\ \mu}(r)
&= -\delta^{ij} \frac{\pi^{2} }{6 }\,C Q^2  r^4 K_1^2(Qr) 
\,,
\\
\Psi_{\ \ \  \mu}^{++\ \mu}(r)
&=   -\frac{ \pi^{2} }{6}\,C s  r^4 K_0^2(Qr)\,,
\end{align*}
where $i,j$
run over the  transverse space  $\mathbb{R}^2$ directions and $K$ is the Bessel function of the second  kind. 
The constant $C$ is determined by the conformal two point function 
\begin{equation*}
\langle j^a(y)j^b(0) \rangle= C \,\frac{y^2\eta^{ab}-2y^ay^b}{(y^2+i\epsilon)^4} \ .
\end{equation*}

By dimensional analysis
the integral over $\bar{r}$  in (\ref{lowxW}) is 
$$
\int \frac{d\bar{r}}{\bar{r}^{2-\omega}}\,\Phi(\bar{r}) = \frac{h(\omega)}{M^{1+\omega}}\,,
$$
where $h(\omega)$ is a dimensionless function that depends on the details of the IR physics associated
to the target hadron wave function but does not affect  the small $x$ scaling behavior of the structure functions  in
the AdS  black disk model. Indeed, taking the imaginary part of
 \begin{align*}
W^{ij}= \delta^{ij}\Pi_1 \ ,\ \ \ \ \ \ \ 
W^{++}\approx \frac{1}{2 x^2}\left( \Pi_2 - {2x} \Pi_1  \right)  \,,
\end{align*}  
equation (\ref{lowxW}) gives
 \begin{align*}
 &F_2-2xF_1\approx
  x^{-\omega} \left(Q/M\right)^{1+\omega}   \frac{ \pi^{\frac{5}{2}} \Gamma^3\left(\frac{3+\omega}{2}\right)C\,h(\omega)}{12\Gamma \left(\frac{4+\omega}{2}\right)}\,,
\\
&F_1\approx
 \left(\frac{Q}{xM}\right)^{1+\omega} 
 \frac{\pi^{\frac{5}{2}} \Gamma \left(\frac{5+\omega}{2}\right) \Gamma \left(\frac{3+\omega}{2}\right) \Gamma \left(\frac{1+\omega}{2}\right)C\,h(\omega)}{24\,\Gamma \left(\frac{4+\omega}{2}\right)}\ .
\end{align*}
Moreover, our ignorance about the target hadron wave function drops out from the ratio $F_L /F_T = \frac{1+\omega}{3+\omega}$. 
We conclude that, in the AdS black disk model and at small $x$, $F_L /F_T$ attains a universal value uniquely fixed by the exponent $\omega$ that controls the growth of the structure functions at small $x$. 
We emphasize that this universal value is independent of the nature of the target hadron. 
Moreover, since the current wave function $\Psi$ localizes in the UV region $r \lsim 1/Q$, our result should be robust against deviations from conformal symmetry in the IR.

The cross section leading to the above structure functions violates the Froissard bound. This is not surprising because we assumed conformal symmetry. In fact,
the same violation occurs in the BK equation \cite{KW}. It is expected that in QCD confinement effects will restore the Froissard bound \cite{IM}. 
Attempts to understand this bound at strong coupling using  the gauge/gravity duality include \cite{Giddings,KN}. 
It would be very interesting to understand this connection in the weak coupling limit.

More complete expressions for the structure functions inside the saturation region can be  easily derived  by integrating each of the terms in (\ref{saturatedW}).
As data for low $x$ DIS inside saturation becomes available from future colliders, it would be very interesting to test the general formula
predicted by the AdS black disk model, as done in \cite{Saturation}.

\section{Photon-photon scattering}
Now we consider scattering of highly virtual photons  \cite{Budnev,gamma-gamma}.
The total cross section can be determined from the imaginary part of the forward scattering amplitude,
\begin{equation*}
(2\pi)^4\, \delta\left( \sum k_j \right)  i\,T^{abcd} =  
\left\langle  j^a(k_1) j^c(k_2) j^b(k_3) j^d(k_4) \right\rangle\,.
\end{equation*}
Using the impact parameter representation and the AdS black disk model we can write
\begin{equation}
T^{abcd} \approx  2\pi i s^{1+\omega}
\int \frac{dr}{r^{2-\omega}} \,
\Psi^{ab\ \mu}_{\ \ \mu}(r)
\int \frac{d\bar{r}}{\bar{r}^{2-\omega}}\,\bar{\Psi}^{cd\ \mu}_{\ \ \mu} (\bar{r})\,,
\label{gammagamma}
\end{equation}
where $\Psi$ is given by (\ref{Psi}) and 
 \begin{align*}
\bar{\Psi}^{cd\ \mu}_{\ \ \mu}(\bar{r})
& = -\frac{    \pi^{2} }{6  }\,C\,  \bar{r}^2  \int_0^\infty dudv \,
e^{-u-v-\frac{\bar{Q}^2 \bar{r}^2}{4u}-\frac{\bar{Q}^2 \bar{r}^2}{4v}}    
\\ & \times\  
 \left(
\begin{array}{ccc}
   \frac{4(u-1)(v-1)}{s\bar{r}^2} &  \frac{v-1}{u}   
  &0\\
  \frac{u-1}{v} & \frac{s \bar{r}^2}{4uv } 
     &  0  \\
  0&0&\mathbb{I}  
 \end{array}
\right)\,,\end{align*}
with $k_1$ and $k_2$ given by (\ref{k1k2}) with $M^2$ replaced by $-\bar{Q}^2$.

To determine the total cross sections we use \cite{Budnev}
\begin{align}
\frac{1}{s}\,{\rm Im}\, T^{abcd}&\approx  R^{ab}R^{cd}\,\sigma^{TT}+R^{ab}P_2^{cd}\,\sigma^{TL}
\label{genform} \\&+
P_1^{ab}R^{cd}\,\sigma^{LT}+P_1^{ab}P_2^{cd}\,\sigma^{LL}\ , \nonumber
\end{align}
where
 \begin{align*}
 R^{ab}&=\eta^{ab}-\frac{k_1\cdot k_2 \left(k_1^a k_2^b+k_2^a k_1^b\right)-k_2^2\, k_1^a k_1^b-k_1^2\, k_1^a k_1^b}
 {(k_1\cdot k_2)^2-k_1^2k_2^2}\,,
\\
 P_1^{ab}&=\frac{\left( k_1\cdot k_2 \,k_1^a-  k_1^2 \, k_2^a\right)  \left(  k_1\cdot k_2 \,k_1^b-  k_1^2  \,k_2^b\right)}{k_1^2\,(k_1\cdot k_2)^2}\,.
\end{align*}
$P_2^{ab}$ is obtained from $P_1^{ab}$ by exchanging  $1\leftrightarrow 2$.  In general, (\ref{genform}) also contains interference terms but these vanish 
for the AdS black disk model. 
At high energies, the total cross section for transverse photons is then
\begin{equation*}
\sigma^{TT}\approx \frac{1}{s}\, f(\omega) \left( \frac{s}{Q\bar{Q}}\right)^{1+\omega}\ ,
 \end{equation*}
where
\begin{equation*}
 f(\omega)=C^2 \frac{\pi^{6} \Gamma^2 \left(\frac{5+\omega}{2}\right) \Gamma^2 \left(\frac{3+\omega}{2}\right) \Gamma^2 \left(\frac{1+\omega}{2}\right) }{72\,\Gamma^2 \left(\frac{4+\omega}{2}\right)}\ .
   \end{equation*}
Similarly, one can determine the total cross section for longitudinal polarized photons. As in DIS, the ratios 
\begin{equation*}
 \frac{\sigma^{LT} }{\sigma^{TT}}\approx \frac{\sigma^{LL} }{\sigma^{LT}}
 \approx  \frac{1+\omega}{3+\omega} 
\ , 
   \end{equation*}
attain universal values for $s \gg Q^2 \sim \bar{Q}^2$. 
Clearly, $(\sigma^{LT} )^2\approx \sigma^{TT}\sigma^{LL} $, independently of $\omega$. 
This equality also holds for an amplitude  given by a single Regge pole (pomeron) and follows from factorization of its residue \cite{GribovPom}. 
Equation (\ref{gammagamma}) shows that a similar factorization occurs in the AdS black disk model at high energy.

In (\ref{gammagamma}), after integrating in the AdS impact parameter, we considered only  the leading term $(sr\bar{r})^{\omega}$. It is easy
to include the other terms similar to those in 
(\ref{saturatedW}), obtaining the more general formulae predicted by
the AdS black disk model for  polarized cross sections inside saturation.

\section{\label{Geometric Scaling} Relation to Geometric Scaling }

We conclude  showing that the growth of the AdS black disk with $S$ given in (\ref{AdSdisk})
can be seen as arising from the exchange of a Reggeon. 
The value of the exponent $\omega$
may be fixed experimentally using geometric scaling  \cite{geometric}. 
The use of a single Reggeon to describe general features
of the crossover from  a dilute system of partons within the hadron to a very dense one is well known \cite{sat,SatSaddle}.
In the context of AdS/CFT, similar arguments 
for external scalar states were given in  \cite{Saturation}, so we shall be brief.

To estimate the AdS black disk size, 
consider the linear term in the expansion of $e^{i\chi}$ in (\ref{forward}),
and  approximate  $\chi$  with the contribution of a single Reggeon exchange \cite{CCP09}
\begin{align}
 \chi^{\mu}_{\   \tau}   & \approx   \frac{1}{N^2}   \int  
  d\nu~ S^{j(\nu)-1}\label{Regge_chi}
\\& \times
\left[ \beta (  \nu)\, \delta^{\mu}_{\ \tau} 
+\tilde{\beta}(\nu) \left( \nabla^{\mu} \nabla_{\tau} - \frac{1}{3} \,  \delta^{\mu}_{\ \tau}\right) \right] \Omega_{i\nu} \left(L \right) \,,
\nonumber
\end{align}
where  $\nabla_{\mu}$ is the Levi-Civita connection on $H_3$ and
$\Omega_{i\nu}(L)=\nu \sin{\nu L}/(4\pi^2\sinh L)$
is a basis of harmonic functions on $H_3$. The Regge spin $j=j(\nu)$  and the residues $\beta(\nu)$ and $\tilde{\beta}(\nu)$ depend
on the 't Hooft coupling $\bar{\alpha}_s=\alpha_s N/\pi$ and are even functions of $\nu$. This is exact in a conformal gauge theory 
such as  $\mathcal{N}=4$ SYM theory and approximate in QCD at weak coupling.

The explicit form of  the Regge residues $\beta(\nu)$ and $\tilde{\beta}(\nu)$ was computed 
to leading order in perturbation theory in \cite{CCP09}. These 
residues are purely imaginary and satisfy ${\rm Im} \,\chi^{\mu}_{\   \tau}  >0$. 
For fixed $S$, the phase shift $\chi(S,L)$  vanishes at large impact parameter $L\rightarrow \infty$. As the impact 
parameter decreases the phase will grow and reach order unity at $L_s(S)$.
For impact parameters $L<L_s$ the amplitude will then be that of an absorptive black disk and the details in the form of the 
phase shift  for $L<L_s$ are not important to the computation of the hadronic tensor in (\ref{forward}). 
To estimate the size $L_s$ of the AdS black disk one can do a saddle point approximation to 
the $\nu$ integral in (\ref{Regge_chi}), exactly as in \cite{Saturation}. At weak coupling the 
Reggeon is the perturbative BFKL hard pomeron \cite{BFKL} and one obtains 
$\omega\approx 2.44 \bar{\alpha}_s$. 
This value of $\omega$ is actually to large to match experiments, and next to leading order correction to the BFKL spin are important.
A similar estimation
of the saturation region can be done using the dipole formalism 
\cite{SatSaddle}, however, we 
remark that here $L_s$ is the black disk size in the AdS impact parameter space $H_3$. 

Saturation effects at strong coupling in DIS have been analysed using the AdS/CFT duality in \cite{HIM}.
We can also show that
at strong coupling and  for a conformal gauge theory, 
the size of the AdS black disk still satisfies  (\ref{AdSdisk}). 
In this case for large impact parameter the phase shift is predominantly real and is given by the AdS gravi-Reggeon trajectory. 
In particular, the residue $\beta(\nu)$ has  poles at $\nu=\pm 2i$, as required for graviton exchange \cite{Lorenzo}, while
$\tilde{\beta}(\nu)$ vanishes \cite{Bartels}. From
conservation of energy-momentum  $j(\pm 2i,\bar{\alpha}_s)=2$ for any value of the coupling, so the $\nu$ integral in 
(\ref{Regge_chi}) can be done by contour deformation. One obtains  $\chi\sim e^{3(L_s-L)}$, with 
$L_s$ given by  (\ref{AdSdisk}) with exponent $\omega=1/3$.
Thus, for $L<L_s$ the phase shift  becomes a rapid varying phase for the $L$
integration in the hadronic tensor (\ref{forward}) and can therefore be dropped.

We finish by relating 
the phenomenon of 
geometric scaling \cite{geometric}, associated to the growth
of the hadron gluon distribution function, to the exponent
$\omega$.  For small $x$  
the cross section depends on a scaling variable $\tau$ defined by
\begin{equation}
\tau = \left( \frac{Q}{Q_s}\right)^2\,,\ \ \ \ \ \ \ \ Q_s^2= M^2 \left(\frac{1}{x} \right)^{\lambda}\,,
\label{GeometricScaling}
\end{equation}
where $Q_s=Q_s(x)$ is the saturation scale. The exponent $\lambda$ is determined experimentally to be 
$\lambda=0.321\pm0.056$ \cite{geometric2}.
Geometric scaling occurs over a large kinematical window that includes the region $Q\sim Q_s$. 
In particular, in this near saturation region,
and provided $Q\gsim 1\,{\rm GeV}$ so that $\alpha_s$ is small and runs slowly,
one can relate the exponents 
$\omega$ and $\lambda$ by associating the growth of the cross section to a single Reggeon exchange.

At  zero momentum transfer the amplitude (\ref{IPrep})  becomes
\begin{equation}
 T^{ab} \approx  2is   \int \frac{dr}{r^3}\, \frac{d\bar{r}}{\bar{r}^3} \,
\Psi^{ab\ \tau}_{\ \ \mu}(r)\, \Phi (\bar{r})  \,  t^{\mu}_{\ \tau} (s,r,\bar{r})
\nonumber
\end{equation}
where, in the linear approximation,
\begin{equation}
t^{\mu}_{\ \tau} (s,r,\bar{r})= \int dl_\perp\, i\chi_{\ \tau}^\mu (S,L) \,.
\nonumber
\end{equation}
Using the Regge representation (\ref{Regge_chi}), we can do
the integration in $l_\perp$ of the harmonic function $\Omega_{i\nu} \left(L \right)$ and its derivatives \cite{CCP09}.
In particular, the components $T^{ij}$ are
\begin{equation}
 T^{ij}  \approx  2is   \int \frac{dr}{r^3}\, \frac{d\bar{r}}{\bar{r}^3} \,
\Psi^{ij\ \mu}_{\ \ \mu}(r)\, \Phi (\bar{r})  \,  t_T (s,r,\bar{r})
\label{LinearT}
\end{equation}
with  
\begin{equation}
t_T= 
 \frac{i}{s}  \int d\nu\, \frac{3\beta(\nu) +(3i\nu-4)\tilde{\beta}(\nu) }{6\pi N^2} \left( sr\bar{r}\right)^{j(\nu) } \left( \frac{\bar{r}}{r} \right)^{-i\nu}\,.
\nonumber
\end{equation}
Since the minimum value of the AdS impact parameter  $L$ is given by $\ln(\bar{r}/r)$, the region near saturation corresponds to 
setting  $\ln(\bar{r}/r)\sim L_s(sr\bar{r})$, which will be the dominant region in the radial integration in (\ref{LinearT}) for a suitable
choice of  the external kinematics $s$ and 
$Q$. In this case the $\nu$ integral for the amplitude $t_T$ is dominated by the saturation saddle point $\nu_s$, and it scales as
$$
t_T \sim 
\frac{1}{s}\, \tau^{-(1+i\nu_s) \frac{1-\omega}{2}}\,,\ \ \ \ \ \ \ \ 
\tau= \frac{\bar{r}^2}{r^2} \left( s r^2\right)^{-\frac{2\omega}{1-\omega}}\,,
$$
where now the scaling variable $\tau=\tau(r,\bar{r})$.
The same scaling holds for the component $T^{++}$ of the amplitude. Integrating over  
$r$ and $\bar{r}$, we conclude that the structure functions $F_T$ and $F_L$ will depend on $\tau$ given in (\ref{GeometricScaling}) with
exponent  $\lambda= 2\omega/(1-\omega)$. From the experimental value of $\lambda$ we may therefore  estimate
$\omega=  0.138\pm 0.021$.
 Using this value, we predict $F_L/F_T \approx 0.36$ deep inside saturation.

\section*{Acknowledgments}
This work was partially supported by grants  CERN/FP/109306/2009,  SFRH/BPD/34052/2006, NSF PHY05-51164 and
by  the FCT POCI  programme.

\end{document}